\documentclass[twoside]{ae100prg}
\usepackage{graphicx}
\usepackage[breaklinks]{hyperref}
\usepackage{booktabs}

\begin{document}
\title{Quantum singularities in conformally static spacetimes }


\author{Deborah A. Konkowski$^1$ and Thomas M. Helliwell$^2$}

\address{$^1$ Mathematics Department, U.S. Naval Academy, Annapolis, Maryland 21402 USA}
\address{$^2$ Physics Department, Harvey Mudd College, Claremont, California 91711 USA}

\email{dak@usna.edu, helliwell@hmc.edu}

\begin{abstract}
After a brief review of the standard definition and analysis of classical singularities in general relativistic spacetimes, and of quantum singularities in static spacetimes with timelike classical singularities,  an extension of quantum singularities to conformally static spacetimes is summarized and applied to two test cases. The timelike classical singularities in a Friedmann-Robertson-Walker (FRW) universe with a cosmic string, and in Roberts spacetime, are shown to be quantum mechanically singular when tested by either minimally coupled or conformally coupled scalar waves. In the Roberts case, however, non-minimally coupled scalar waves with a coupling constant $\xi \ge 2$ do not detect the classical singularity. 

\end{abstract}

\section{Introduction}
\par We study quantum wave packet propagation in conformally static spacetimes with timelike classical singularities. If the wave propagation turns out to be well defined, the spacetimes are said to be quantum mechanically non-singular. 

\par The order of the paper is as follows: First, classical and quantum singularities are defined with the latter restricted (as usual) to static spacetimes with timelike singularities. Next, the definition of quantum singularity is extended to conformally static spacetimes with a timelike singularity (spacelike singularities, if present,  are not tested). In particular, two spacetimes are tested with generally coupled scalar waves: a Friedmann-Robertson-Walker (FRW) spacetime with a cosmic string and the Roberts spacetime. Finally, conclusions are given, together with ideas for further research.

\section{Classical singularities}

\par A spacetime $(M, g)$ is taken to be a paracompact, $C^{\infty}$, connected, Hausdorff manifold $M$ with a Lorentzian metric $g$ \cite{ES}. So what is a classical \emph{singularity}? A spacetime is by definition smooth, so `singular' points are not part of the spacetime; they must be cut out of the spacetime manifold. This leaves a `hole', with incomplete curves, a seeming boundary to spacetime.  How do we complete spacetime, and how do we define a boundary $\partial M$ to spacetime?  There have been a number of attempts, none of them entirely satisfactory. Note that Cauchy completeness works only in Riemannian metrics, not Lorentzian.  Boundary definitions have included the a(abstract)-boundary of Scott and Szekeres \cite{SS}, the b(bundle)-boundary of Schmidt \cite{S}, the c(causal)- boundary of Geroch, Kronheimer, and Penrose \cite{GKP} and the g(geodesic)-boundary of Geroch \cite{G}. In this discussion we will use Geroch's 1968 description of a classical singularity. He states that ``a singularity is indicated by incomplete geodesics or incomplete curves of bounded acceleration in a maximal spacetime.'' This is closest to the definition of classical singularity used in the famous singularity theorems of Hawking and Penrose, which predict that singularities are  ubiquitous in exact solutions of Einstein's equations (see, e.g., \cite{HE})). 

\par Ellis and Schmidt have classified singular points into three types according to their strength \cite{ES}: quasi-regular (mild, topological singularities), non-scalar curvature (diverging tidal forces on curves ending at the singularity; finite tidal forces on some nearby curves) and scalar curvature (diverging scalars -- usually one considers only $C^0$ scalar polynomial invariants). Conical singularities, as in idealized cosmic strings, are a good example of quasiregular singularities. The other two types of singularities are stronger, curvature singularities.  Nonscalar curvature singularities include those in whimper cosmologies and certain plane-wave spacetimes, whereas  scalar curvature singularities are the best-known, occurring at the centers of black holes or the beginning of big bang cosmologies.

\subsection{Quantum Singularities}

What happens if instead of classical particle paths (e.g., null and timelike geodesics) one uses quantum mechanical particles (quantum wave packets) to identify singularities? Following pioneering work by Wald \cite{wald}, Horowitz and Marolf answered this question for static spacetimes with timelike classical singularities. In their 1995 paper they posit that a spacetime is quantum mechanically (QM) \emph{non}singular if the evolution of a test scalar wave packet, representing a quantum particle, is uniquely determined by the initial wave packet, the manifold and the metric, without having to place boundary conditions at the classical singularity. Technically, a static spacetime is QM-singular if the spatial portion of the Klein-Gordon operator is not essentially self-adjoint on $C_{o}^{\infty}(\Sigma)$ in  the space of square integrable functions $L^{2}(\Sigma)$, where $\Sigma$ is a spatial hypersurface. 

\par The term ``essentially self adjoint"  arises in functional analysis \cite{RS}. An operator $A$ is called self-adjoint if (i) $A = A^*$ and (ii) $Dom(A)$ = $Dom(A^*)$, where $A^*$ is the adjoint of $A$ and $Dom$ is short for domain. An operator is \emph{essentially} self-adjoint if (i) is met and (ii) can be met by expanding the domain of the operator $A$ or its adjoint $A^*$ so that it is true. 

There are two basic tests for essential self-adjointness \cite{RS}. The first uses the von Neumann criterion of deficiency indices \cite{vN}; one studies solutions of $A \Psi = \pm i \Psi$, where $A$ is the spatial portion of the Klein-Gordon operator, and finds the number of solutions for each sign that are self-adjoint. The second technique uses the so-called Weyl limit point - limit circle criterion \cite{weyl}, which relates essential self-adjointness of the Hamiltonian operator to the behavior of the ``potential'' in an effective one-dimensional Schr\"odinger equation, which in turn determines the behavior of the scalar wave packet. Relevant theorems that simplify the analysis can be found in Reed and Simon \cite{RS}. 

\par Many authors have used the definition of quantum singularity to study the singularity structure of spacetimes. For a summary, see, for example, the review article by Pitelli and Letelier \cite{PL} or the conference proceeding by the authors \cite{KH} and the references therein. Also, there is the alternative concept of `wave regularity'  introduced by Ishibashi and Hosoya \cite{IH}, which is relevant to the discussion. It uses a non-standard Hilbert space, $H^{1}$, the first Sobolev space.

\section{Conformally Static Spacetimes}

 A spacetime $g_{\mu\nu} (x^\alpha )$ that is conformally static is related to a static spacetime $\bar{g}_{\mu\nu}( x^a)$ by a conformal transformation $C(\eta)$ of the metric.  Here $C(\eta)$ is the conformal factor, where $\eta$ is the conformal time, related to the time $t$ by $dt = C d\eta$.  Simply put,  $ g_{\mu\nu}(x^\alpha) = C^2(\eta) \bar{g}_{\mu\nu}(x^a)$. Here Greek letters $\alpha, \beta, ...$ label spacetime indices and have the range over 0, 1, 2, 3, and Latin letters $a, b, c, ...$ label spatial indicies that range over 1, 2, 3.

\par The Lagrangian density for a generally coupled scalar field is \cite{BD},

\begin{equation}
\mathcal{L} = 1/2 (-g)^{1/2} [ g^{\mu \nu} \Phi,_{\mu} \Phi,_{\nu} - (M^{2} + \xi R)\Phi^{2}], 
\end{equation}

\noindent where $M$ is the mass if the scalar particle, $R$ is the scalar curvature, and $\xi$ is the coupling (in particular, $\xi=0$ for minimal coupling and $\xi=1/6$ for conformal coupling). Varying the action $ S = \int \mathcal{L} \ d^{4}x$ gives the Klein-Gordon field equation,

\begin{equation}
|g|^{-1/2}\left(|g|^{1/2}g^{\mu \nu} \Phi,_{\nu}\right),_{\mu} - \xi R\Phi=M^2\Phi.
\end{equation}.

\noindent In the massless case with conformal coupling, the field equation above is conformally invariant under a conformal transformation of the metric and field; in this case the inner product respecting the stress tensor for the field is also conformally invariant. This led Ishibashi and Hosoya to state \cite{IH}, in the case of wave regularity, that ``the calculation is as simple as that in the static case when singularities in conformally static space-times are probed with conformally coupled scalar fields.''

\par Here we study the quantum particle propagation in spacetimes with massive scalar particles described by the Klein-Gordon equation and the limit point - limit circle criterion of Weyl \cite{weyl} \cite{RS}. In particular, after separating variables we study the radial equation in a one-dimensional Schr\"odinger form with a `potential' and determine the number of solutions that are square integrable. If we obtain a unique solution, without placing boundary conditions at the location of the classical singularity, we can say that the solution to the full Klein-Gordon equation is quantum mechanically (QM) nonsingular. The results depend on the spacetime metric parameters and wave equation modes.

\par After separating variables we take the spatial portion to be an operator equation on a Hilbert space $L^{2}(\Sigma)$ with inner product (see, e.g., \cite{K}),

\begin{equation}
(\chi, \zeta) = \int d^3 x |\bar g_3/g_{00}|^{1/2} \chi(x^a) \zeta(x^b), 
\end{equation} 

\noindent where $\bar g_3$ is the determinant of the spatial portion of the static metric, $\chi$ and $\zeta$ are spatial mode solutions and $a,b$ range over 1, 2, 3.  Then we consider the radial portion alone, change variables and write the radial equation in one-dimensional Schrodinger form, $H u(x) = E u(x)$,  where the operator $H = -d^2/dx^2 + V(x)$ and $E$ is a constant, with the singularity at $x=0$. The inner product here is simply $\int dx |u(x)|^2$ and the Hilbert space is $L^2(0,\infty)$. At this point one can simply apply the limit point - limit circle criterion as easily as in the static case in order to determine the quantum singularity structure.

\subsection{FRW with a Cosmic String}

A simple metric modeling a Friedmann-Robertson-Walker cosmology with a cosmic string \cite{DS} is given by

\begin{equation}
ds^2= a^2(t)( -dt^2 + dr^2 + \beta^2 r^2 d\phi^2 +dz^2)
\end{equation}

\noindent where $\beta=1-4\mu$ and $\mu$ is the mass per unit length of the cosmic string. This metric is conformally static (actually conformally flat).  Classically it has a scalar curvature singularity when $a(t)$ is zero and a quasiregular singularity when $\beta^2\neq1$. Here we will consider the timelike quasiregular singularity alone. The Klein-Gordon equation with general coupling can be separated into mode solutions

\begin{equation}
\Phi = T(t) H(r) e^{im\phi} e^{ikz}
\end{equation}

\noindent where

\begin{equation}
\ddot{T} + 2 \left( \frac{\dot{a}}{a}\right) \dot{T} + (M^2 a^2 + \xi R a^2 - q) T =0
\end{equation}

\noindent and

\begin{equation}
H'' + \frac{1}{r} H' + (-k^2 - q - \frac{m^2}{\beta^2 r^2}) H = 0.
\end{equation}

\noindent The $T$-equation alone contains $M$ and $R$. Rewriting the dependent and independent variables as  $r=x$ and $ H = x u(x)$, we get the correct inner product form and a one-dimensional Schr\"odinger equation,

\begin{equation}
u'' + (E - V(x))u = 0
\end{equation}

\noindent where $E =-k^2 - q$ and

\begin{equation}
V(x) = \frac{m^2 - \beta^2 /4}{\beta^2 x^2}.
\end{equation}.

\noindent Near zero one can show that the potential $V(x)$ is limit point if $m^2/\beta^2 \geq 1$. Therefore any modes with sufficiently large $m$ are limit point, but  $m=0$ is limit circle; thus generically this conformally static space-time is quantum mechanically singular.

\subsection{Roberts Spacetime}

The Roberts metric \cite{Roberts} is

\begin{equation}
ds^2 = e^{2t}(-dt^2 + dr^2 + G^2(r) d\Omega^2)
\end{equation}

\noindent where $G^2(r) = (1/4)[ 1+ p - (1 -p) e^{-2r}]( e^{2r} - 1)$. The spacetime is conformally static,  spherically symmetric, and self-similar (see, e.g., \cite{IH}).  It has a timelike classical scalar curvature singularity at $r = 0$ for $0 < p < 1$. The Klein-Gordon equation can be solved by separation of variables with mode solutions given by $\Phi= T(t) H(r) Y_{lm}(\theta, \phi)$. The radial operator can be put in one-dimensional Schr\"odinger form and the limit point - limit circle criterion applied. Details are given in \cite{HK}. One finds that the spacetime is quantum mechanically singular if $\xi <  2$ and quantum mechanically non-singular if $\xi \ge 2$. Therefore, the classical timelike singularity remains singular when probed by minimally coupled ($\xi = 0$) waves or by conformally coupled ($\xi = 1/6$) waves.

\section{Conclusions}
After a brief review of the standard definition and analysis of classical singularities in general relativistic spacetimes, and of quantum singularities in static spacetimes with timelike classical singularities,  an extension of quantum singularites to conformally static spacetimes was summarized and applied to two test cases. The timelike classical singularities in a FRW universe with a cosmic string and in Roberts spacetime were shown to be quantum mechanically singular when tested by either minimally coupled or conformally coupled scalar waves. In the Roberts case, however, non-minimally coupled scalar waves with a coupling constant $\xi \ge 2$ did not detect the classical singularity. 

\par Further analysis of the singularity structure of conformally static spacetimes is underway \cite{HK}. A class of spherically symmetric conformally static spacetimes is being analyzed; this class includes the spacetimes of HMN\cite{HMN} and Fonarev\cite{F}, as well as the Roberts spacetime.

\section*{Acknowledgements} 
One of us (DAK) thanks B. Yaptinchay for useful discussions. 

\section*{References}

\bibliography{konkowski-references}

\end{document}